# Machine Learning Supports Existence of Previously Unrecognized Transient Astronomical Phenomena in Historical Observatory Images

Stephen Bruehl[a], Brian Doherty[b], Alina Streblyanska[c], and Beatriz Villarroel[d]

[a]Department of Anesthesiology, Vanderbilt University Medical Center, 701 Medical Arts Building, 1211 Twenty-First Avenue South, Nashville, TN 37212, USA

[b]Independent Researcher, Dallas, TX, USA

[c]Instituto de Astrofísica de Canarias, Avda Vía Láctea S/N, La Laguna, E-38205, Tenerife, Spain

[d]Nordita, KTH Royal Institute of Technology and Stockholm University, Hannes Alfvéns väg 12, 106 91, Stockholm, Sweden

CORRESPONDING AUTHOR: Stephen Bruehl, Ph.D., Vanderbilt University Medical Center, 701 Medical Arts Building, 1211 Twenty-First Avenue South, Nashville, TN 37212, USA.

E-Mail: Stephen.Bruehl@vumc.org

Competing Interest Statement:

None of the authors have competing interests to declare.

Abstract

Transient, star-like point sources that appear and vanish over short timescales are described in astronomical images prior to launch of Sputnik. We have reported that transient numbers diminish significantly in Earth's shadow (shadow deficit) and are more likely within (+/-) one day of nuclear testing (nuclear window). These findings remain debated with some arguing that transients identified via existing automated pipelines are simply plate defects. Therefore, we used machine learning (ML) to enhance transient identification accuracy and validate the phenomenon. The model was trained against 250 transient image pairs taken 30 minutes apart that were classified as real versus plate defect by expert visual review; the model demonstrated good discrimination (out-of-fold AUC=0.81; sensitivity=0.71, specificity=0.71). After deployment in a dataset of 107,875 previously-identified transients, the model assigned each a probability of being real. After controlling for ML-identified artifacts, transient counts were significantly elevated for dates within a nuclear window ($p = .024$); transients with the highest probability of being real were more likely to occur within a nuclear window ($p<.0001$). The shadow deficit was significant ($p<.0001$) and largest in the highest probability transients relative to lower probability transients ($p=.003$). Results strongly support existence of an unrecognized population of transient objects in historical astronomical plates warranting further study.

KEYWORDS: Transient, Machine Learning, Validation, Nuclear Testing, Shadow Deficit, POSS-I, Astronomical Images

## Introduction

Transient, star-like phenomena exhibiting point spread functions have been identified by comparing sequential images over short timescales in the Palomar Observatory Sky Survey (POSS-I) and other historical sky surveys[2,8,9,11-14]. Their morphology suggests they may persist on sub-second timescales (i.e., brief flashes)[15]. These short-lived transients are absent in images taken immediately before the transients appear and in all images from subsequent surveys[8,9,11,12] (see Figure 1 for example). They do not correspond with known objects in optical catalogs (e.g. Gaia, Pan-STARRS) and cannot be accounted for by known satellites or orbital launch debris, as all images were obtained prior to the launch of the first artificial satellite[9,12].

The nature and origin of transients are unclear and difficult to elucidate due to their historical nature. Modern attempts to capture similar phenomena are fraught due to the quantity of satellites and launch debris currently in orbit. Nonetheless, recent findings point to potentially intriguing transient characteristics. Transient numbers have been reported to diminish by ~30% when located within Earth's shadow, suggesting a “shadow deficit” consistent with some transients representing highly reflective objects in orbit[5,14]. Furthermore, transients are significantly increased when within a “nuclear testing window” (+/- one day of historical above-ground nuclear weapons tests) relative to days with no nuclear testing[1,3,5]. Such contemporaneous correlates of transients potentially provide information for better characterizing their nature.

Validity of these findings has been debated based largely on the potentially high (and unknown) level of false positives in existing transient data given their automated identification on scanned photographic plates more than 50 years old likely prone to plate defects (e.g., emulsion errors, dust, scratches)[6,16,17]. Both the shadow deficit and nuclear testing association were identified in a dataset of 107,875 transients developed by Solano et al. using an automated pipeline[9], with no visual inspection to document validity of individual transients identified (which was not feasible due to the extensive work required).

To address the unknown false positive rate in transient identification using existing automated methods, the current study used machine learning (ML), specifically a supervised learning approach, to quantify the probability that each transient of the n = 107,875 previously identified reflected a real object versus plate defect. To validate the resulting ML model against external criteria, we tested two *a priori* hypotheses based on our prior findings[1,14]. Both were predicated on the assumption that if transients represent real phenomena, decreased statistical noise due to reduction of false positives in the ML-vetted dataset would enhance signal to noise ratio and increase effect sizes. Specifically, we hypothesized that both the magnitude of the shadow deficit reported in Villarroel et al.[14] and the strength of the transient-nuclear testing association reported in Bruehl & Villarroel[1] would be largest in the subgroup of transients that the ML model determined to be most likely to be real.

## Methods

### Data Sources

#### *Transient Data*.

The transient dataset to which ML was applied consisted of 107,875 transients occurring between 11/19/49 and 4/28/57 identified in publicly-available scanned images from the POSS-I survey available on the DSS Plate Finder website (https://archive.stsci.edu/cgi-bin/dss_plate_finder). The process used initially to identify these transients and eliminate misidentifications was conducted via an automated workflow detailed in Solano et al[9] and further clarified in Villarroel et al.[16]. In brief, transients were defined as distinct star-like point sources exhibiting point spread functions that were present in POSS-I Red images but that were absent both in images taken immediately prior to the POSS-I Red image and in all subsequent images from other surveys. A final criterion for classifying an object as a transient was that there were no optical counterparts either in PanStarrs DR1 or Gaia DR3 at less than 5 arcsec (see Villarroel et al.[14] for additional details).

*Nuclear Weapons Testing Data*.

Replication of the nuclear testing-transient association reported in Bruehl & Villarroel[1] employed public data sources. Dates of all U.S. above ground nuclear tests during the study period noted above were obtained from the US Department of Energy (DOE/NV-209 Rev.16). As in our prior work[1], the nuclear testing variable was again a nuclear testing window reflecting whether each date fell within one day of any nuclear weapons test (test date +/- 1 day). This window was chosen to balance sensitivity to the effects of interest with a reasonable degree of specificity, given numerous anecdotal reports of unidentified aerial phenomena both during and shortly before or after historic nuclear tests[7].

Procedure

*Machine Learning Model Development*.

Machine learning (ML) model development was conducted in Python 3.11 using the scikit-learn (v1.7.2), XGBoost (v3.1.2), and SHAP (v0.50.0) libraries. Transients targeted for model development in the current study were all identified solely on red POSS-I plates. The ML model included 23 predictors extracted from the red FITS images and the VASCO v4 catalog. Seven catalog-level morphometric features were included: signal-to-noise ratio (SNR), point spread function (PSF) ratio, elongation, compactness, sharpness, number of comparison stars, and candidate score (described in Solano et al.[9]). The ML model also included 6 plate-aggregate features: a plate quality indicator, plate-level SNR fraction, SNR standard deviation, mean elongation, and high-SNR source count. Finally, the model included 10 morphometric features identified by the ML model in the red FITS images themselves: PSF Full Width at Half Maximum (PSF FWHM), ellipticity, sharpness, connected pixel count, aperture flux, distance to plate edge, symmetry score, gradient magnitude, proximity to bright star, and FITS-measured SNR. The ML model did not include any spectral features or red-blue plate comparison.

The criterion for training and determining accuracy of the ML model was a training set of 250 POSS-I image pairs (red versus blue) taken 30 minutes apart and previously identified by the Solano et al. pipeline as containing at least one transient on the red image[9,16]. The training set was chosen to include exemplars of either true astronomical transients or a variety of clearly-identifiable plate defects (e.g., dust, hair, scratch, emulsion defect, exposure problems). Each image pair was manually inspected by an astronomer with expertise regarding the transient phenomenon (BV); 134 were labelled likely real transients and 116 as plate defects. To enhance classification consistency over time, a second reviewer (SB) who was trained by the primary evaluator periodically reviewed a subset of the images assigned to the primary reviewer, with any disagreements discussed with the primary reviewer and resolved.

The ensemble ML classifier detailed in this study combined four tree-based models (XGBoost, Random Forest, Gradient Boosting, LightGBM), each trained with 300 trees and identical hyperparameters, with final classification predictions based on the unweighted mean of the four models' predicted probabilities. For analytic purposes, transients were distributed into 10 bins of equal size (probability deciles) with each increasing decile reflecting a higher probability that transients in that decile reflect real transients based on the ML model (see Table 1). Isotonic probability calibration was attempted but collapsed the probability distribution; raw mean ensemble probabilities were therefore used as the primary candidate probability scores described in analyses (i.e., likelihood that each transient would be classified as real). Additional details regarding the ML methods can be provided upon reasonable request to the authors. The final machine learning script can be found at: https://github.com/dca-doherty/VASCO-ML.

*Determination of the Shadow Deficit.*

Transients appear as point sources rather than streaks during a 50-minute plate exposure, and therefore calculation of the shadow deficit assumed that transients were at geostationary orbit altitude, 35,786 km (GEO). In our prior work[14], we employed the Nir et al. 2D EarthShadow model (https://github.com/guynir42/earthshadow) to calculate the shadow deficit. In the current work, we used a 3D model in order to more accurately capture the hypothesized shadow deficit. Specific advantages of the 3D model were that it enabled the shadow boundary to be computed correctly for any assumed orbital altitude, it permitted anchoring of the model at Mount Palomar's actual coordinates, and it accounted for the impact of Earth's penumbra on shadow deficit results.

Each transient's angular separation (based on recorded sky position) from the center of Earth's penumbra was computed using the topocentric anti-solar direction as seen from the coordinates of Palomar Observatory, with candidate coordinates precessed from j2000.0 to the observation epoch using the IAU 1976 formulae. The primary model was the 3D topocentric penumbra, which has an angular radius of 8.96 degrees at GEO and represents the boundary at which solar illumination is fully restored. Transients falling within this angular radius were classified as in-shadow; all others were classified as sunlit.

The fraction of retained transient candidates falling within the shadow zone was compared to the fraction expected to appear based on geometric expectations for a uniform sky distribution. The latter plate-aware control sample was generated as follows. Individual plates were identified by spatial clustering of transient positions per observation date (complete linkage, 15-degree threshold), yielding 626 plates. For each plate, 100 random sky positions were generated within the plate field of view using the plate center coordinates and observation time, with no reference to actual transient positions. Of 62,600 total control positions, 433 (0.692%) fell within Earth's geometric shadow at geosynchronous orbit altitude, establishing the expected shadow fraction of 0.692%. The shadow deficit for each decile of probability values (i.e., increasingly stringent definitions of a transient) was calculated as: (observed shadow fraction – control shadow fraction) / control shadow fraction, expressed as a percentage. See Villarroel et al.[14] for additional details.

Statistical Analysis

All statistical analyses were conducted in Python 3.11 using the statsmodels, scipy, and scikit-learn libraries. To demonstrate the discrimination accuracy of the ML model, we examined sensitivity (True Positive / True Positive + False Negative), specificity (True Negative / True Negative + False Positive), and area under the curve (AUC), all based on 5-fold cross-validation with final results comprising cross-validated out-of-fold values. Sensitivity and specificity were also derived from pooled out-of-fold values. All tests of ML model discrimination used manually-classified transients as described above as the reference standard. Given that both study hypotheses were *a priori* and based on our prior findings, we used one-tailed probability values to determine statistical significance in all analyses described below.

*Analysis of the Shadow Deficit Hypothesis*

The shadow deficit quantifies extent to which transients are underrepresented in Earth's shadow relative to geometric expectations. The shadow deficit is presented in Figure 3 for consistency with our prior work[14] but all analyses focused solely on observed shadow fraction values. For each probability decile, the observed fraction of transients falling within Earth's shadow was compared to a plate-aware Monte Carlo control expectation of 0.692% using a one-sided binomial test. Because ten simultaneous tests were conducted, a Bonferroni-corrected significance threshold of $p < 0.005$ was applied. To test whether the highest-probability transients showed a disproportionately larger shadow deficit compared to lower probability transients, the shadow fraction of the upper decile (D9) was compared to the pooled shadow fraction of all other deciles (D0 through D8) using a two-proportion z-test.

*Analysis of the Transient-Nuclear Testing Hypothesis*

Primary analyses of hypothesized transient-nuclear testing associations were restricted in this study to U.S. nuclear tests, which were most often conducted at the Nevada test site, a location only 435 kilometers from Mount Palomar Observatory (in California) from which all POSS-I transients were identified. In contrast, the bulk of remaining worldwide nuclear tests were conducted by the Soviet Union in Semipalatinsk, Kazakhstan, which is 10,617 kilometers from Mount Palomar. This focus on U.S. nuclear testing in the current study was intended to address questions raised in response to our prior work[1] regarding mechanistic plausibility of nuclear-transient associations if nuclear tests were conducted vast distances from the observatory from which transients were identified.

The nuclear-transient association was tested in primary analyses using a non-parametric permutation approach (i.e., no distributional assumptions) with observation dates as the independent unit of analysis ($n = 370$ dates of POSS-I observations). This focus of analyses on dates as the unit of analysis avoided possible confounding of results using other available analytic approaches due to temporal clustering of transients (non-independence within dates).

For each date, the sum of ML-assigned probabilities across all transients on that date was computed, yielding a probability-weighted transient count that represents the estimated number of real transients. The mean probability-weighted count for nuclear dates (those falling within one day of a US nuclear test, 26 of 370 dates) was compared to the comparable count on non-nuclear dates. Statistical significance was assessed via 10,000 random permutations of the nuclear window labels (i.e, the label indicating whether each date was within a nuclear window). This analytic approach therefore tested how often randomly-shuffled nuclear window label assignments produce a mean difference for summed probability values as large as what was actually observed. A confirmatory non-parametric Mann-Whitney U test was also conducted given that it has more statistical power for detecting shifts within continuous distributions. To parallel the shadow deficit analysis above, we compared the percentage of transients in a nuclear testing window for the upper decile (D9) to the pooled percentage of transients in a nuclear testing window for all other deciles (D0 through D8) using a two-proportion z-test.

To assess the temporal specificity of nuclear-transient associations identified, we repeated the primary permutation test approach above for different lagged time periods, from 3 days before a nuclear test date to 3 days after a nuclear test date. This sensitivity analysis tested whether the nuclear-transient signal was sharply localized around test dates or broadly distributed, which may better distinguish a genuine physical association from confounding by seasonal patterns or observation scheduling. For purposes of comparing the size of the nuclear-transient association between different lags for significant analyses, we report the unadjusted ratio of the mean probability-weighted transient count on nuclear-window dates to the mean on non-nuclear dates.

## Results

### Machine Learning Model Accuracy

The training set for ML-based supervised learning comprised 250 image pairs (red versus blue POSS-I images taken 30 minutes apart) classified as containing at least one transient appearing only on the red plate based on the Solano et al. automated pipeline[9,16]. This training set was manually inspected by an astronomer with expertise in the transient phenomenon (BV); 134 were labelled likely real transients and 116 as plate defects. The ML model included 23 morphometric and other predictors extracted solely from the red plate (see Methods for ML details). No spectral features or red-blue plate comparisons were included in the model. Shapley (SHAP) values quantifying relative contributions of each predictor to the final ML model are displayed in Figure 2.

The final ML model demonstrated good overall discrimination relative to manual labelling, with an area under the curve (AUC) value of 0.81 +/- 0.04 across 5-fold cross-validation. The sensitivity was 0.71 (probability of being classified as a real transient if manual inspection categorized the transient as likely real) and specificity was 0.71 (probability of being classified as a plate defect if manual inspection classified the transient as a defect). The ML model was then deployed in the original n = 107,875 transient dataset that was analyzed in our prior work[1,14], assigning each transient a probability of being labelled a real transient. All transients were assigned to 10 equal-sized bins (deciles) of increasingly greater ML-derived probability that the transients contained in each bin reflected real transients (lowest probability transients were in Decile 0 and the highest probability transients were in Decile 9).

Table 1 summarizes transient probabilities by deciles. Of the transients in the original dataset, only the upper 20 % (Deciles 8 and 9) exceeded a 0.66 probability of being a real transient. Only 10% of the original sample (Decile 9) approached or exceeded at least a 0.80 probability of being a real transient.

While the majority of transients occurred in isolation, multiple transients were not uncommon. When considering only the highest confidence group of transients (Decile 9) within 10 arcmin of each other on the same night of observation, we found that there were 549 doublets (transient pairs), 143 triplets, and 190 larger groupings.

Validation of the Machine Learning Model

*Model Sensitivity to Hypothesized Shadow Deficit.*

As described in Villarroel et al.[14], the number of transients in Earth's shadow was expected to decrease substantially relative to sunlit regions to the extent that transients represent orbital objects exhibiting specular reflections. If the ML model were successfully removing plate defects and increasing the relative signal of real transients, the shadow deficit previously reported[14] was expected to be largest in the subset of transients with the highest ML-derived probability scores (Decile 9). That is, transients most likely to be real would show the largest shadow deficit, with this physical property validating both the transient phenomenon and the ML model. Details of the shadow deficit calculation methodology are described in the Methods and in Villarroel et al.[14]. In the current work, shadow deficit analyses compared the observed number of transients in Earth's shadow (the shadow fraction) to the number theoretically-expected given a random geometric distribution (0.692% expected; see Methods).

Figure 3 presents the observed shadow deficit across the 10 ML-derived probability deciles. Note that to facilitate interpretation in context of the current hypotheses and prior work, Figure 3 displays the shadow *deficit*, reflecting the difference between the observed shadow fraction and the expected random distribution shadow fraction. However, all related analyses focus on the shadow fraction itself as the dependent variable.

A series of binomial tests was conducted to compare the observed shadow fraction for each decile to the expected shadow fraction for a random geometric distribution. Results of these binomial tests for Deciles 1, 5, 6, and 9 exceeded the Bonferroni-corrected significance threshold of $p < 0.005$, indicating that there was a significant shadow deficit compared to expectations for transients assigned to these probability deciles. However, the smallest shadow fraction (0.31%), corresponding to the largest shadow deficit (55.2%) compared to expectations, was observed in the decile with the highest probability of reflecting real transients based on the ML model (Decile 9); $p<.001$), a finding significant at the 5 sigma level. Given the current study hypotheses, it is notable that the shadow fraction in Decile 9 was significantly smaller than the pooled mean of all other deciles (two-proportion z test = 2.78, $p = 0.003$). Moreover, follow-up Bonferroni-corrected pairwise tests between the highest probability decile (Decile 9) and the two deciles with the lowest probability of being real transients (Decile 1 and Decile 2) confirmed that the shadow fraction was significantly smaller for the highest confidence transients (exceeding the Bonferroni corrected significance threshold of $p = 0.0056$). These results all support our findings as indicating that the largest shadow deficit occurred in the transient subgroup that the ML model determined to have the highest probability of reflecting real transients.

*Model Sensitivity to Hypothesized Transient Associations with Nuclear Testing.*

Of the 370 POSS-I observation nights, 26 fell within a United States (U.S.) nuclear weapons testing window (+/- 1 day of a nuclear test; see Methods for rationale). This definition was adopted for consistency with our prior related work[1]. In the current work, we tested the hypothesis that if the ML model were successfully eliminating false positive transients, the strength of association between nuclear testing and number of transients would increase as the ML-derived probability that transients were real increased (i.e., across increasing probability deciles). The percentage of transients observed within a nuclear window is presented by probability decile in Figure 4. Inspection of this figure suggests that the association between transients and nuclear testing strengthens linearly as the probability that transients are real increases. There was a 62.7% higher rate of transients occurring within a nuclear window for the highest probability decile (Decile 9; 13.6%) relative to the lowest probability decile (Decile 0; 8.3%). The percentage of transients in Decile 9 that fell within a nuclear window significantly exceeded the pooled mean percentage falling within a nuclear window across the remaining nine deciles (11.1%; two-proportion z-test, $z = 7.40$, $p < 0.0001$). The latter statistical findings should be considered descriptive only rather than hypothesis testing due to potential confounds that are addressed in the primary analyses below.

In primary analyses, we tested the nuclear-transient hypothesis using a non-parametric permutation testing approach (i.e., no distributional assumptions) with observation dates as the independent unit of analysis. This focus on dates as the unit of analysis avoided possible confounding of results inherent in several other potential analytic approaches that might be biased by temporal clustering of transients (non-independence within dates). For each date, the sum of ML-assigned probabilities across all transients on that date was computed, yielding a probability-weighted transient count that represented the estimated number of real transients on that date. Each permutation test evaluated whether the summed probability values for dates within a nuclear window differed significantly from the summed probability value expected for a random distribution of nuclear labels (i.e., the label indicating whether there was a nuclear test on that date) based on 10,000 permutations. See the Methods for the rationale and additional details. This permutation test confirmed that probability-weighted transient counts were significantly higher for dates within a nuclear window compared to the value expected based on a random distribution ($p = 0.024$). A follow-up Mann-Whitney U test, which also makes no distributional assumptions but is more powerful statistically, confirmed that probability-weighted transient counts were significantly higher for dates within a nuclear window compared to non-nuclear dates ($p = 0.002$). These findings indicate that even when the influence of artifacts in the original transient sample is controlled for, there are significantly more transients recorded on dates within a nuclear window.

Post-hoc sensitivity analyses were conducted to examine the temporal specificity of the nuclear-transient association across all probability deciles. This analysis could not be restricted only to the highest probability transients due to sample size limitations. We examined a series of lags from 3 days before to 3 days after each nuclear test date. These analyses were conducted to determine whether the apparent nuclear-transient signal is sharply localized around test dates or broadly distributed, with a highly specific signal providing further support for the effect representing a genuine physical association rather than a spurious correlation (e.g., due to observation scheduling). A series of permutation tests modelled after the primary analyses were conducted for the series of lags noted above. The probability-weighted transient count (summed probabilities for each day) was significantly higher than expected compared to 10,000 random distributions for both the day of the nuclear test ($p = .0355$) and the lagged timing one day before the nuclear test ($p = .0288$). The lagged permutation test that was focused on the day after the nuclear test also approached significance ($p = 0.078$). The magnitude of these effects can be expressed as the ratio of probability-weighted mean transient count for the nuclear testing time frame being targeted to the comparable transient count for days with no nuclear testing. The largest effect was one day before the nuclear test (ratio = 2.57), with analyses on the day of the nuclear test having only a slightly smaller ratio of 2.36. For comparison, this ratio of weighted probability scores for nuclear test dates to non-test dates was 1.70 for the analysis of the day after a nuclear test (which approached significance), but ranged from 0.31 to 1.37 for all other lagged tests (all of which were nonsignificant, p's>0.19). The pattern of findings in these sensitivity analyses indicated that the nuclear-transient association was highly specific temporally, consistent with it reflecting a real association. We note that the lagged finding for one day prior to a nuclear test which exhibited the largest effect must be interpreted cautiously, as the POSS-I plates were exposed overnight whereas the nuclear tests were generally conducted the following morning. Thus, the "lag" one day before a nuclear test actually reflects transients that were present closest in time to the nuclear test itself.

## Discussion

Transients are short-lived, star-like objects of unknown origin that have been identified in the POSS-I and other historical sky surveys[2,4,8,9,11-14]. A significant barrier to accepting transients as real astronomical phenomena and to understanding their nature and origin is the unknown and potentially high rate of false positive transient identifications using the existing published automated pipeline[9,16]. These false positives comprise plate defects due to dust, hair, scratches, and emulsion errors on scanned archival astronomical plates. This study for the first time attempted to reduce these false positives and optimize identification of true astronomical transients via a supervised learning ML approach.

Results indicated the ML model displayed good discrimination in terms of correspondence with classifications as real transient versus plate defect based on expert visual inspection (out-of-sample AUC = 0.81). Some have argued that all purported transients are simply plate defects[6,17]. Our results clearly argue against this possibility. If this were true, the supervised learning approach we employed would have trained the ML model simply to recognize plate defects that had all been improperly, and presumably randomly, labelled as true transients versus plate defects. In this scenario, ML model discrimination would be expected to approximate random levels (AUC = 0.50), which contrasts sharply with the observed AUC = 0.81.

Inspection of the range of probability values for each of the 10 increasingly stringent deciles do indicate that the original n = 107,875 dataset from the Solano et al. pipeline[9,16] examined in our prior work[1,14] contains a high proportion of false positives. Only 20% of transients in this dataset exceeded a 0.66 probability of being a real transient. Only 10% of this original sample (Decile 9) approached or exceeded at least an 0.80 probability of being a real transient. Overall, only a small subset of the original n = 107,875 dataset were judged by the ML model to have high likelihood of reflecting a real object rather than a plate defect, highlighting the issue of statistical noise in the existing pipeline as well as the potential value of incorporating this ML model into future efforts to better understand transient characteristics.

Given the model's success at removing false positives and the expected improvement in signal-to-noise ratio in the ML-vetted transient data, we proceeded to evaluate two *a priori* hypotheses to validate the model against external criteria. First, Villarroel et al.[14] had reported that the number of transients was significantly lower (~7.6 sigma) when located in Earth's shadow compared to expectations given actual sky coverage (i.e., shadow deficit). In the current study, we found that the magnitude of this shadow deficit was largest in the subgroup in which it would be expected according to our hypotheses, that is, the subgroup with the highest probability of reflecting real objects according to ML model. The shadow fraction in this group was significantly lower than the mean across all other groups, indicating that the shadow deficit magnitude was greatest in this high probability subgroup. While the magnitude of the shadow deficit in the current work cannot be directly compared with findings in our prior related work[14] due to use of different Earth shadow models (3-D penumbral model in the current work), our findings nonetheless provide additional support for the previously proposed hypothesis that a subset of transients exhibit features consistent with those expected of highly reflective objects in orbit[5,8].

Our second means of validating the ML model was to test the *a priori* hypothesis that associations between nuclear testing and transients reported in our prior work[1] and confirmed by others[3,5] would be strongest in the subgroup of transients the ML model judged most likely to be real objects rather than plate defects. Results confirmed this hypothesis as well. The probability-weighted transient count was significantly higher for dates within a nuclear test window than the transient count expected based on a random distribution (reflecting 10,000 permutations). By directly factoring the ML-derived probability scores into the transient count for each day, this analytic approach demonstrated that transient counts were significantly higher than expected for days within a nuclear window even when controlling for any influence of the high percentage of likely plate artifacts in the sample. We also highlight that sensitivity analyses indicate the association between greater transient counts and nuclear testing was specific to the day of the nuclear test and the day immediately prior (when the transient images closest in time to the actual test were obtained). There were no other significant nuclear-transient associations noted from 3 days prior to each nuclear test through 3 days after each test. The temporal specificity of the nuclear-transient association and the fact that this association is statistically significant even when plate artifacts are controlled both confirm and strengthen our prior findings[1].

Taken together, results of this study argue strongly against claims that all transients are plate defects. If this were true, the magnitude of both the shadow deficit and the nuclear-transient association would not have shown any systematic association with ML-derived probability values. The fact that the shadow deficit and the nuclear-transient association were each significantly larger in transients with the highest ML-derived probability of being real supports the transient phenomenon, the shadow deficit, and the nuclear-transient association all as reflecting real physical phenomena rather than being due to spurious associations arising from plate defects.

So how should the two key findings in the current study be interpreted? A number of potential explanations for each of these findings *when considered in isolation* have been addressed previously[1,14,16]. It is not difficult to accept that the shadow deficit finding, considered on its own, might imply that there is an unknown population of previously undiscovered objects in orbit with high specular reflectivity and flat surfaces. But how can this interpretation account for the finding that this same group of high-probability transients is observed more often around the dates of nuclear weapons testing? Does one simply disregard the latter finding because it does not fit within current scientifically-accepted models? If one accepts that both key findings in this study represent real phenomena and their interpretation is necessarily limited to hypotheses that can simultaneously account for both findings, we are left with a very small set of provocative hypotheses that would be considered highly implausible by most. For example, is it possible that unknown to the public there were multiple launches of artificial satellites long before Sputnik with some launches timed to coincide with U.S. nuclear tests? Or rather, do the current findings represent detection of a non-human technosignature? Due to data limitations, such hypotheses cannot be subjected to falsification. Ultimately, it must be left to the reader to determine the most plausible explanation for the current findings. Regardless, we note that replication of our findings using scanned images from sky surveys obtained at other observatories worldwide during the same time period (1949-1957) could at minimum independently corroborate the validity of the associations reported here. More work in this regard needs to be carried out.

Prior critiques had previously disputed interpretation of transients identified in historical astronomical images as representing an unrecognized population of real transient astronomical phenomena distinguishable from plate defects[6,17]. However, together with several recent replications of our prior work[3,5], the current findings provide compelling data that argue strongly for considering high probability transients as real astronomical phenomena worthy of further study. A broad methodological implication of our study findings is that ML can be successfully used to extract higher quality astronomical data from historical sky surveys that often are underutilized. For example, having sequential images as in POSS-I of the same sky region over relatively short timescales (30 minutes apart) that have in effect been digitally cleaned via ML could potentially enhance their utility for time domain astronomy, such as detection of asteroids and Kuiper belt objects, monitoring novae, or characterizing pulsars and M-dwarf stellar flares.

The current study addresses several limitations of our prior work. Comments by critics who have dismissed the transient phenomenon as simply being plate defects[6,17] have been addressed by developing and validating an ML approach permitting quantitative discrimination between real transients and plate defects via ML-derived probability scores. Comments that previously reported associations between presence/absence of nuclear testing and presence/absence of transients[1] were simply due to coincidental correspondence between POSS-I observation schedules and nuclear testing schedules (potentially reflecting common weather patterns)[17] were addressed by: 1) limiting nuclear-transient analyses only to recorded dates of POSS-I observation (n = 370) with no assumptions regarding transient presence/absence on other dates, and 2) targeting nightly transient *counts* as the variable of interest, which unlike presence/absence of transients, should not exhibit spurious association with nuclear testing schedules or common weather patterns influencing both observing conditions and nuclear test schedules. Finally, the current work used a 3D topocentric model for calculating shadow deficits rather than the 2D model used in our initial work[14], given that the 3D model permitted more accurate estimation of the shadow deficit (see Methods).

Although this study has several strengths, there are also limitations. The primary limitation was the reliance upon subjective expert opinion to determine the classification of training sample images as real transients versus plate defects. This was unavoidable due to the absence of any objective gold standard for determining the validity of a given transient image. While there is potential for classification error with this approach, this situation parallels challenges routinely encountered in research seeking to validate medical diagnostic criteria. For example, in developing the widely used Diagnostic and Statistical Manual of Mental Disorders III-Revised, validation research employed expert clinical opinion regarding diagnosis as the gold standard[10,18]. This approach was necessary given the absence of clear biomarkers for the disorders targeted, a situation paralleling the transient classification issues in the current work. We note that the individual primarily responsible for performing the manual inspection of transient images (BV) is the co-discoverer of the phenomenon and therefore well-qualified to make classification decisions. Moreover, a separate individual (SB) trained by the primary reviewer periodically reviewed a subset of training images to serve as a check on classification consistency of the primary reviewer, with any disagreements resolved by discussion.

In conclusion, our study strongly suggests that a population of previously unknown transient objects were detected on photographic plates by historical astronomical surveys although their nature remains unexplained. Results confirm that a ML approach can be used to optimize discrimination of real transients from plate defects. Findings that transients classified by the ML model as likely real exhibit the largest shadow deficit and the strongest associations with nuclear testing both validate the ML model and the transient phenomenon itself. What these two findings taken together might reveal about the nature of the transient phenomenon remains to be confirmed, although additional research on this topic is clearly warranted.

## Acknowledgments

B.V. is funded by the Swedish Research Council (Vetenskapsr\aa det, grant no. 2024-04708) and supported by a generous donor. A.S. is supported by the Athanatos Foundation.  The authors would like to express their appreciation to Dr. Enrique Solano for his work creating the original transient identification pipeline on which the current study was based.

Table 1. Characteristics of each ML-based probability decile derived from the original n = 107,875 transient dataset generated by the Solano et al.[9] pipeline.

| Probability Decile | N | Probability Range | N Candidates Retained | Cumulative % of Full Catalog |
|---|---|---|---|---|
| D0 | 10,788 | 0.013 - 0.072 | 107,875 | 100% |
| D1 | 10,787 | 0.072 - 0.105 | 97,087 | 90% |
| D2 | 10,788 | 0.105 - 0.150 | 86,300 | 80% |
| D3 | 10,787 | 0.150 - 0.215 | 75,512 | 70% |
| D4 | 10,788 | 0.215 - 0.304 | 64,725 | 60% |
| D5 | 10,787 | 0.304 - 0.410 | 53,937 | 50% |
| D6 | 10,787 | 0.410 - 0.532 | 43,150 | 40% |
| D7 | 10,788 | 0.532 - 0.663 | 32,363 | 30% |
| D8 | 10,787 | 0.663 - 0.780 | 21,575 | 20% |
| D9 | 10,788 | 0.780 - 0.973 | 10,788 | 10% |

Figure 1. The left image presents examples of three transients (green circles) identified on the POSS-I red plate taken on August 11, 1953 (RA = 339.8697437 and Dec = 38.4411717). The image at right is the same region in a POSS-II red image taken on September 3, 1989. The yellow circle indicates a plate defect on the left image.

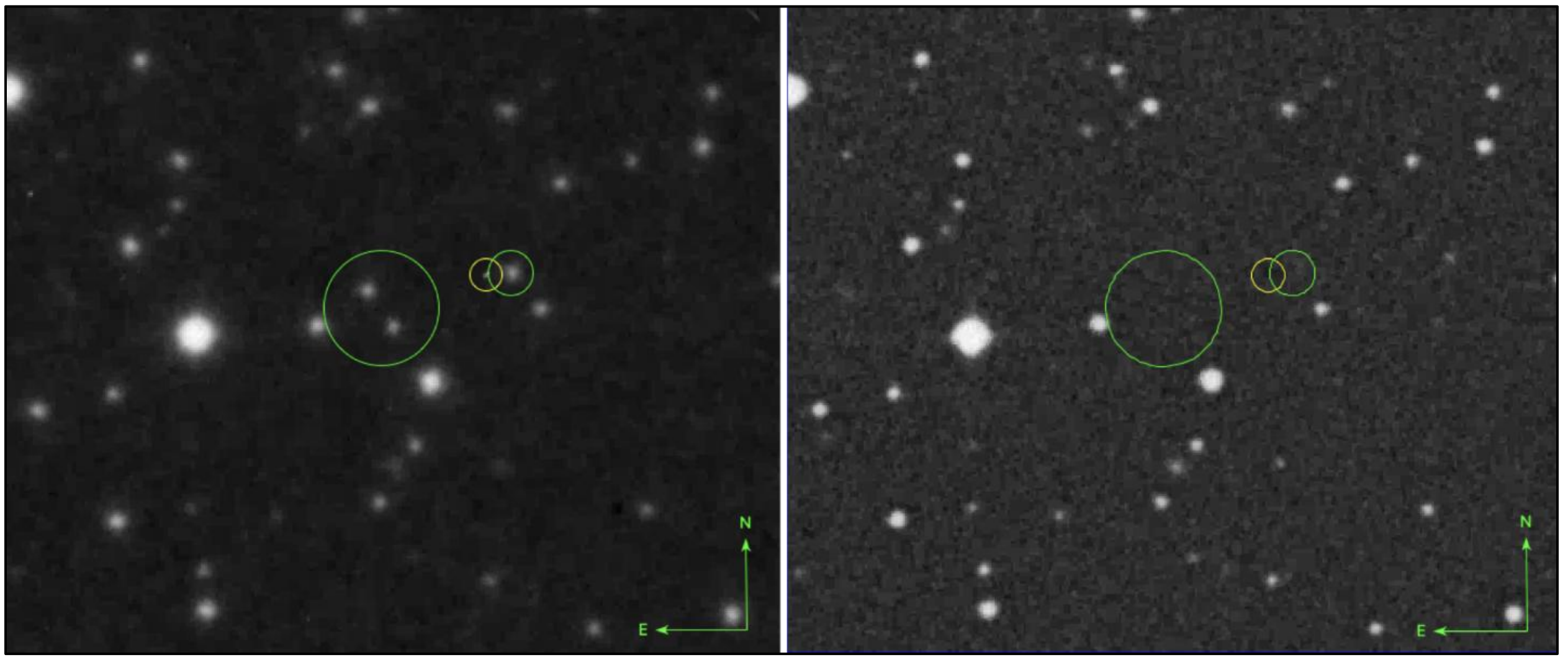

Figure 2. SHAP values for the 23 machine learning predictors of real transients versus plate defects relative to expert classification based on manual inspection of 250 transient candidates from the original n = 107,875 transient dataset. SHAP values indicate the relative contribution of each predictor to the model.

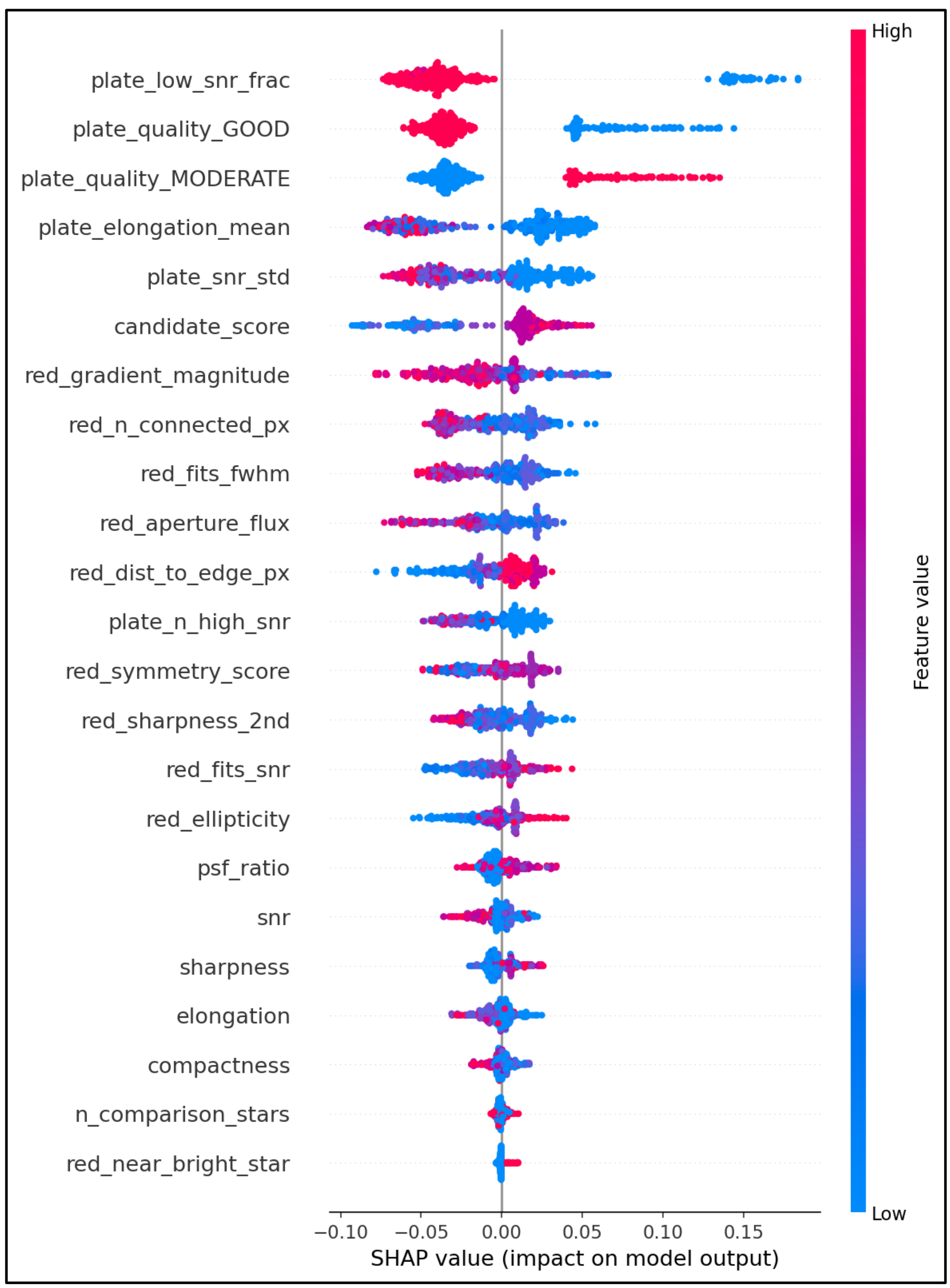

Figure 2 Footnote: Features are listed in descending order of mean absolute SHAP value. Predictor abbreviations are as follows: plate_low_snr_frac = Plate-level: fraction of candidates on the same plate with SNR < 5; plate_quality_GOOD = Plate-level: binary, plate rated as GOOD (one-hot encoded); plate_quality_MODERATE = Plate-level: binary, plate rated as MODERATE (one-hot encoded); plate_elongation_mean = Plate-level: mean elongation of all candidates on the same plate; plate_snr_std = Plate-level: standard deviation of SNR across all candidates on the same plate; candidate_score = Candidate-level: original pipeline quality score from Solano et al. detection algorithm; red_gradient_magnitude = Red FITS: magnitude of the intensity gradient at candidate position (edge sharpness); red_n_connected_px = Red FITS: number of contiguous pixels above threshold in the source footprint; red_fits_fwhm = Red FITS: full width at half maximum of the candidate PSF profile (point-source width); red_aperture_flux = Red FITS: total flux within the measurement aperture centered on the candidate; red_dist_to_edge_px = Red FITS: distance from candidate to nearest plate edge, in pixels; plate_n_high_snr = Plate-level: count of candidates on the same plate with SNR > 15; red_symmetry_score = Red FITS: radial symmetry of the candidate light profile (1 = perfect symmetry); red_sharpness_2nd = Red FITS: second-order sharpness (curvature of the PSF peak; point sources vs extended); red_fits_snr = Red FITS: signal-to-noise ratio measured directly from the FITS pixel data; red_ellipticity = Red FITS: ellipticity of the source (0 = circular, 1 = elongated); psf_ratio = Candidate-level: ratio of candidate FWHM to plate-median FWHM (PSF consistency check); snr = Candidate-level: signal-to-noise ratio from the original detection pipeline (catalog-level); sharpness = Candidate-level: sharpness from the original detection pipeline; elongation = Candidate-level: elongation from the original detection pipeline; compactness = Candidate-level: compactness from the original detection pipeline; n_comparison_stars = Candidate-level: number of comparison stars used for PSF calibration on that plate; red_near_bright_star = Red FITS: binary flag, candidate is near a bright cataloged star (halo/diffraction risk).

Figure 3. Magnitude of the shadow deficit across ML-based probability deciles. The range of probability values within each decile is indicated in parentheses below each. Increasing deciles reflect greater probability that the transients within each decile reflect real objects rather than plate defects. *** $p < 0.001$, ** $p < 0.01$, * $p < 0.05$ for one-sided binomial tests comparing the observed shadow fraction in each decile against the plate-aware Monte Carlo expected rate of 0.692%.

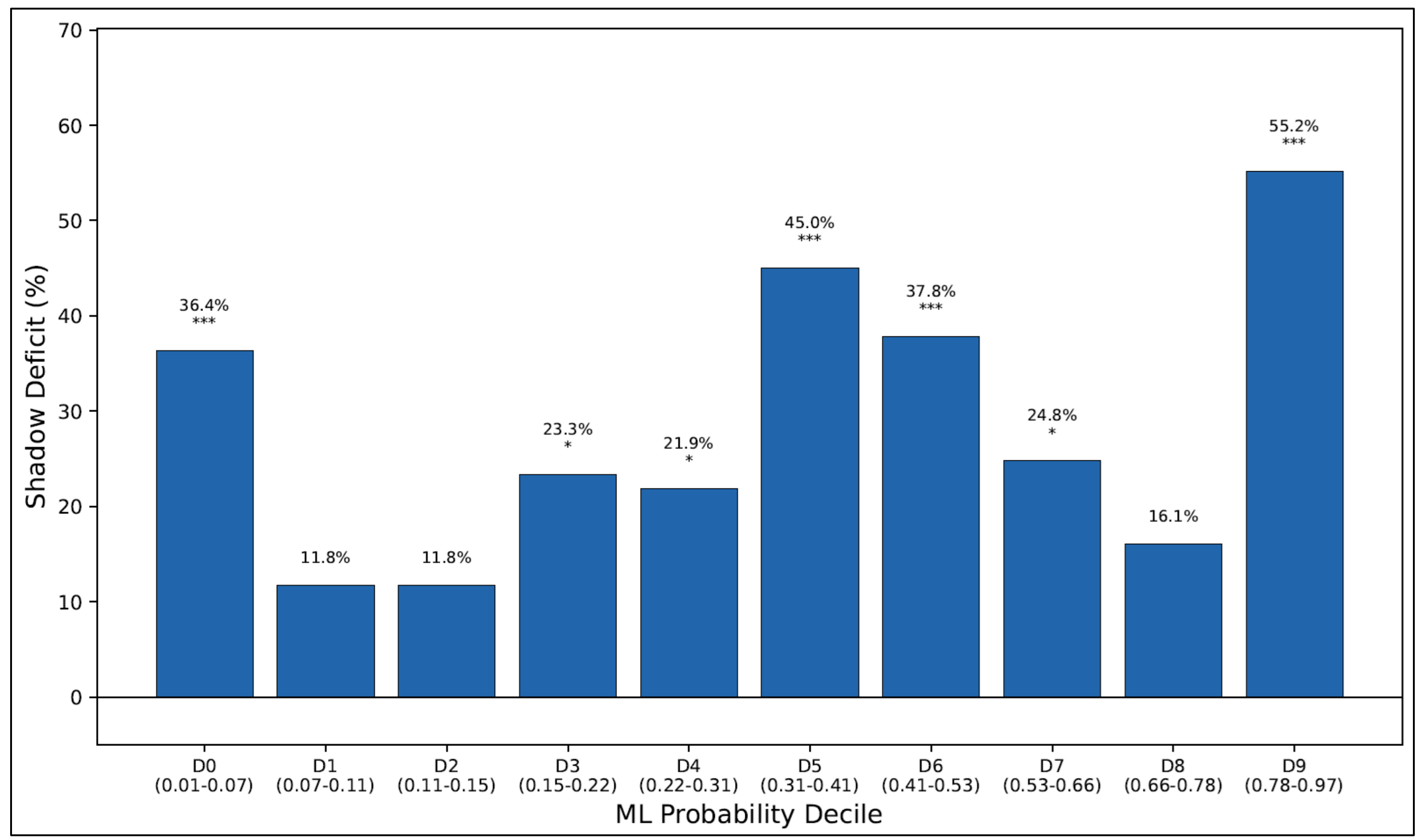

Figure 4. Percentage of transients assigned to each ML-based probability decile that occurred on dates falling within a nuclear window (+/- 1 day of a nuclear weapons test). The sample reflects n=370 days of POSS-I observations.

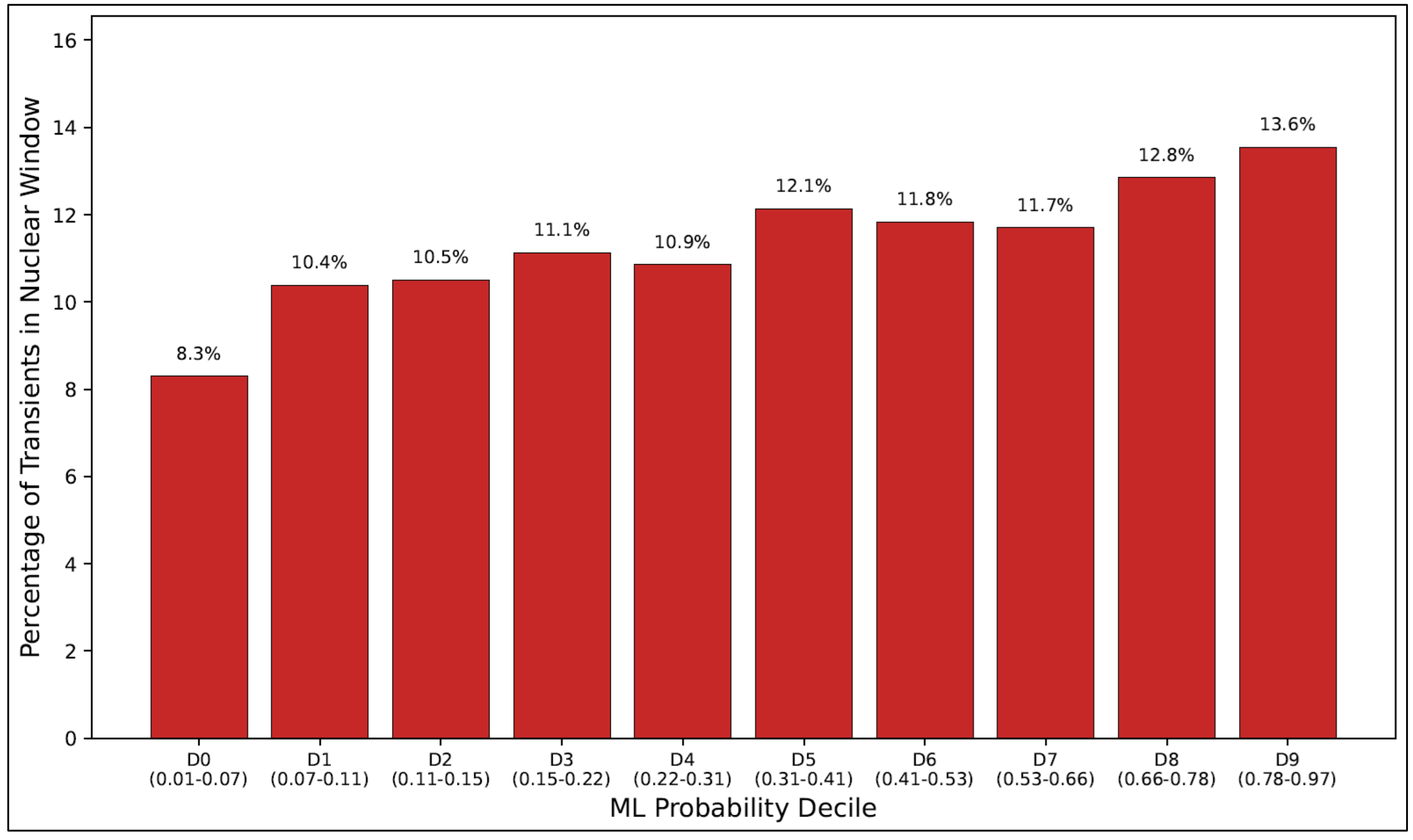